# Identification of the K-most Vulnerable Entities in a Smart Grid System

Sohini Roy, Arunabha Sen
School of Computing, Informatics and Decision System Engineering
Arizona State University
Tempe-85281, Arizona, USA
Email: {sohini.roy, asen}@asu.edu

*Abstract*—A smart grid system can be considered as a multi-layered network with power network in one layer and communication network in the other. The entities in both the layers exhibit complex intra-and-interdependencies between them. A reliable decision making by the smart grid operator is contingent upon correct analysis of such dependencies between its entities and also on accurate identification of the most critical entities in the system. The Modified Implicative Interdependency Model (MIIM) [1] successfully captures such dependencies using multi-valued Boolean Logic based equations called Interdependency Relations (IDRs) after most of the existing models made failed attempts in doing that. In this paper, for any given integer K, this model is used to identify the K-most vulnerable entities in a smart grid, failure of which can maximize the network damage. Owing to the problem being NP complete, an Integer Linear Programming (ILP) based solution is given here. Validation of the model [1] and the results of the ILP based solution is done by simulating a smart grid system of IEEE 14-Bus using MATPOWER and Java Network Simulator (JNS). Simulation results prove that not only the model MIIM [1] is correct but also it can predict the network damage for failure of K-most vulnerable entities more accurately than its predecessor Implicative Interdependency Model (IIM) [2].

*Keywords—smart grid, cascading failure, vulnerable entities, interdependency relations, joint network, MIIM model.*

## I. INTRODUCTION

The smart grid system can be modelled as a joint power-communication network where the entities of both the network are structurally and functionally dependent on each other. The entities of communication network get power supplies from the power network entities and they in turn take the responsibility of monitoring the power network entities by continuously sending SCADA and PMU data from each of the substations to the control centers in a secure and efficient manner. As a result, entities of both power and communication network exhibit complex intra-and-interdependencies between them where the failure of one or more entities can lead to subsequent failure of multiple other entities leading to a catastrophe. In order to avoid such a condition, the researchers should have a clear understanding of such complex dependencies between the entities and based on that they should be able to identify the most critical entities in the smart grid system, failure of which can maximize the network damage.

Efficient hardening techniques [3] followed for such critical entities can save the smart grid from a huge damage. Yet, in order to identify the most vulnerable entities in the system, clear understanding of the design of the joint power-communication system as well as an appropriate interdependency model to capture the complex dependencies in a smart grid are necessary. Therefore, several attempts are made by the researchers to identify those dependencies and reflect them in a graph-based models [4], [5], yet most of them lack physical realism. A simple and easy Boolean Logic based dependency model, termed as the Implicative Interdependency Model (IIM) is proposed in [2], that accurately captures the complex interdependencies between the entities in a critical infrastructure. However, this model ignores the effect of failure of communication network entities and does not provide a clear picture of the communication network design in a smart grid system. Moreover, the IIM model has a binary nature. Entities in that model can either remain operational or become non-operational, but there is no concept of reduced operability which is a very common feature of the smart grid entities. In [6], a rough idea about the communication network for smart grid is provided but it also lacks the detailed description of the design of communication infrastructure. A Modified Implicative Interdependency Model (MIIM) proposed in [1] takes into account all such drawbacks and provides a clear picture of the smart grid network design by taking inputs from a utility in the U.S. Southwest. It also considers different operational levels of the entities and model the complex dependencies using multi-valued Boolean Logic based equations called Interdependency Relations (IDRs) where the entities in the joint network are considered as Boolean variables, values of which can change between 0 indicating no-operation, 1 indicating reduced operation and 2 indicating full operation. Just by solving these IDRs the operators can identify the operational states of different entities in the network after some initial failure has taken place in the system and thereby recognize the most vulnerable entities in the network.

Even after identifying all the vulnerable entities in the system, the smart grid operator can have a budget constraint of hardening only K entities of the network, where K can be any integer. In that case, it is important to identify the K-most critical entities in the system. The problem of identifying the K-most vulnerable entities in a joint power-communication network is already proved to be NP complete in [2]. Therefore, an Integer Linear Programming based solution for the problem is given in this paper using the MIIM IDRs. Finally, a validation of the results obtained from the proposed method is done by co-simulating the two layers of the smart grid network of IEEE 14-Bus system using MATPOWER and Java Network Simulator (JNS) and the simulation results are compared with that obtained using ILP solutions of both MIIM [1] and IIM [2] IDRs to show that the modified version of IIM is more realistic.

The rest of the paper is structured as follows. Section II gives an overview of the IIM and MIIM models. Problem definition of the self-updating K-contingency list is given in Section III and an ILP based solution to the problem is proposed in Section IV of the paper. A comparative analysis between the K-contingency lists obtained using MIIM, IIM and smart grid network layers co-simulation is provided in Section V. Section VI concludes the paper.

## II. OVERVIEW OF IIM AND MIIM

In both IIM [2] and MIIM [1], the smart grid system can be viewed as a multilayer network, represented as a set $J(E, F(E))$, where $E = P \cup C \cup CP$ which means all entities belonging to power layer, communication layer and the intermediate layer joining power and communication entities; and $F(E)$ represents the set of IDRs. The entities in power layer (layer 1) are considered as P type entities where $P = \{P_1, P_2, \ldots P_n\}$ and entities in ICT layer (layer 2) are named as C type entities where $C = \{C_1, C_2, \ldots C_m\}$. The set $F(E)$ is used in both the models to capture the dependencies among interacting entities in the network. Yet, only structural dependencies are considered to generate the IDRs in IIM and both structural as well as operational aspects of the entities are taken into account while formulating IDRs for MIIM. IIM has a binary nature and the entities in that model can either be operational with a state value of 0 or be non-operational with a state value of 1. The most common feature of reduced operability in critical infrastructures is ignored in IIM. The entities in MIIM can take a value of 0, 1 and 2 indicating no-operation, reduced operation and full operation respectively.

Let $C_i$, an entity of layer 2, be operational if (i) $C_j$ which is another entity of layer 2 and $P_a$ which is an entity of layer 1, are operational, or (ii) $C_k$ which is an entity of layer 2 and $P_b$ which is an entity of layer 1 are operational, and (iii) $C_l$ which is an entity in layer 2 is operational. Then the corresponding IIM IDR for $C_i$ would be: $C_i \leftarrow \left(\left(C_j \cdot P_a\right) + \left(C_k \cdot P_b\right)\right) \cdot C_l$. In this IDR, '.' denotes logical AND operation and '+' denotes logical OR operation. Similarly, the IDR for a P type entity can be expressed.

In MIIM, three Boolean operators are used while formulating the IDRs. The first operator is min-AND, denoted by '○', which selects the lowest of its input values. The second operator is max-OR, denoted by '●', which selects the highest of its input values. The third operator is new_XOR, which is denoted by '◉'. If all the inputs of new_XOR are same, then the output is also same as the inputs. In all other cases the output is 1. This new_XOR operator actually denotes the level of operation of an entity. The truth table for all the 3 new operators are given in Table I.

TABLE I. TRUTH TABLE FOR MIIM OPERATORS

| Input 1 | Input 2 | min-AND | max-OR | new_XOR |
|---------|---------|---------|--------|---------|
| 2 | 2 | 2 | 2 | 2 |
| 2 | 1 | 1 | 2 | 1 |
| 2 | 0 | 0 | 2 | 1 |
| 1 | 2 | 1 | 2 | 1 |
| 1 | 1 | 1 | 1 | 1 |
| 1 | 0 | 0 | 1 | 1 |
| 0 | 2 | 0 | 2 | 1 |
| 0 | 1 | 0 | 1 | 1 |
| 0 | 0 | 0 | 0 | 0 |

In order to illustrate MIIM, let us assume that if an entity in condition (i) or (ii) fails, $C_i$ will still work full operability, but if (iii) is not satisfied then $C_i$ will operate at a reduced level; this relation can be expressed using MIIM IDRs as: $C_i \leftarrow \left(\left(C_j \circ P_a\right) \bullet \left(C_k \circ P_b\right)\right) \circledcirc C_l$. To differentiate between the two models in terms of smart grid system application, the failure of entity $C_l$ for the above IIM and MIIM IDRs are considered and the outcomes are observed in Table II.

TABLE II. OPERATIONAL DIFFERENCE BETWEEN IIM AND MIIM

|        | IIM | MIIM |
|--------|-----|------|
| **STEP 1** | $C_l \rightarrow 0$ | $C_l \rightarrow 0$ |
| **STEP 2** | $C_i \leftarrow (((2.2) + (2.2)).0)$ | $C_i \leftarrow (((2 \circ 2) \bullet (2 \circ 2)) \circledcirc 0)$ |
| **STEP 3** | $C_i \leftarrow ((2 + 2).0)$ | $C_i \leftarrow ((2 \bullet 2) \circledcirc 0)$ |
| **STEP 4** | $C_i \leftarrow (2.0)$ | $C_i \leftarrow (2 \circledcirc 0)$ |
| **STEP 5** | $C_i \leftarrow 0$ | $C_i \leftarrow 1$ |

It is observed in Table II, that for same kind of dependencies, failure of the entity $C_l$ results in the failure of entity $C_i$ in case of IIM but it only reduces the operation level in case of MIIM.

## III. K-MOST VULNERABLE ENTITIES (KVE) PROBLEM

It is important for the operator of a smart grid to identify the most vulnerable entities in the network, even before any kind of failure or damage takes place in the system. An automated system offering identification of KVE in the steady state of a smart grid will help the operator to decide which of the entities in the system should be hardened [3], so that in any case the maximum damage in the smart grid can be avoided. When one or more entities fail in the smart grid system, many other entities fail as a result and this is called cascading failures, and this often might lead to a catastrophe if not arrested in time. This cascade stops when the system reaches a steady state once again. Given an integer K and a power-communication network at a steady state, this problem returns the set of K-most critical entities in the joint network, failure of which can lead to the maximum total number of failed entities in the system at the end of the cascade propagation. It is to be noted that a cascade can only propagate in one direction since an already failed entity cannot be affected again by the cascading failure. Therefore, upper bound of the cascade is $|EG| - 1$; where EG is the total number of edges in the network. A formal definition of the KVE problem using the MIIM [1] model is as follows:

### A. Inputs to the Problem

- (a) A joint network $J(E, F(E))$; where $E = P \cup C \cup CP$
    - $P = B \cup T \cup Batt$ (Buses, Transmission Lines/Transformers, Batteries)
    - $C = SE \cup SRE \cup DRE$ (Substation Entities, SONET-Ring Entities, DWDM-Ring Entities)

o $CP = L \cup R \cup U$ (Power supply lines, RTUs and PMUs)
- (b) Two positive integers K and S

## B. Decision version of the Problem

Does there exist a set of K entities in E whose failure at time t would result in a failure of at least S entities in total at the end of the cascading process?

## C. Optimization version of the Problem

Compute the set of K entities in the joint network $J(E, F(E))$ whose failure at time t would maximize the number of entities failed or in other words minimize the overall system state values at the steady state of cascade propagation.

However, the problem of finding K-most vulnerable entities is NP complete which is already proved in [2]. Therefore, an ILP based solution for the problem is given in section 5 of this paper. Also, validation of the results should be done by comparing the ILP based solution results with the simulation results.

## IV. INTEGER LINEAR PROGRAMMING BASED SOLUTION

In this section, an Integer Linear Programming (ILP) based solution for the KVE problem stated in section 4 of this paper is given. The variable list for the problem is given below–

### A. Variable List

For each entity $e_i \in E$ a variable set $x_{i,t} \forall t, 0 \leq t \leq |E|-1$ is created. The value of $x_{i,t}$ is 2 if it is fully operational, 1 if it is operating at a reduced level of operation and 0 if it is non-operational.

### B. Objective function

The objective function for the problem can be defined as:
$$min \sum_{i=1}^{|E|} x_{i,|E|-1}$$

This implies that, the problem aims at minimizing the system states for all the entities in the smart gird.

### C. Constraint Set

*1) Constraint set 1:* $\sum_{i=1}^{|E|} x_{i,0} = K$, entities failed at time step 0 is K.

*2) Constraint set 2:* $x_{i,d} \leq x_{i,t-1}, \forall t, 1 \leq t \leq |E|-1$. This implies that, an entity can only have a system state value at a time $t > d$, less than or equal to the system state value it had at time d.

*3) Constraint set 3:* Based on the 3 new Boolean operations adopted by MIIM, IDRs can have the following format: $e_a \leftarrow (e_b \odot e_c) \circ (e_m \bullet e_n)$.

*a) Step 1:* Firstly, the above IDR can be reformed as: $e_a \leftarrow z_{bcmn}$ where the new variable $z_{bcmn}$ can be expressed as: $z_{bcmn} \leftarrow (g_{bc}) \circ (h_{mn})$ where the two new variables $g_{bc}$ and $h_{mn}$ can be further represented as: $g_{bc} \leftarrow e_b \odot e_c$ and $h_{mn} \leftarrow e_m \bullet e_n$.

*b) Step 2:* Now, a linear constraint is developed for the z type variable (associated with min_AND operator). In order to evaluate the IDR: $z_{bcmn} \leftarrow (g_{bc}) \circ (h_{mn})$, $z_{bcmn}$ can be represented as: $z_{bcmn} \leq g_{bc,t-1}$ and $z_{bcmn} \leq h_{mn,t-1}, \forall t, 1 \leq t \leq |E|-1$.

*c) Step 3:* A linear constraint is also developed for the h type variable (associated with max_OR operator). In order to evaluate the IDR: $h_{mn} \leftarrow e_m \bullet e_n$, $h_{pq}$ can be represented as: $h_{mn} \geq x_{m,t-1}$ and $h_{mn} \geq x_{n,t-1}, \forall t, 1 \leq t \leq |E|-1$.

*d) Step 4:* For the g type variable, associated with the new_XOR operator, the following linear constraint is developed. The IDR: $g_{bc} \leftarrow e_b \odot e_c$ is represented by the following set of linear equations: $g_{bc} \geq 0$, $g_{bc} \leq max\_state$, where max_state denotes the state value at the highest level of operability for an entity ( 2 in this case), and $N \times g_{bc} \leq x_{b,t-1} + x_{c,t-1}, \forall t, 1 \leq t \leq |E|-1$. Here N is the number of operands on which the new_XOR operator operates.

## V. COMPARATIVE ANALYSIS BETWEEN IIM, MIIM AND SIMULATION RESULTS

In order to simulate the smart grid system with IEEE-14 Bus as the power network and a synthetic yet realistic communication network designed as per the principles proposed in [1], two different simulation platforms are selected. MATPOWER is selected for the simulation of the power layer whereas Java Network Simulator (JNS) is selected for the simulation of the communication layer. An event-driven synchronization [7] between these two kinds of simulation platforms is followed, in which whenever an entity in the power network fails at time $t = t'$, not only the power entities associated with that entity are updated but also all the communication entities receiving power from only that entity are removed before the starting of the next round of simulation at $t = (t' + 1)$. When a communication entity fails, it does not have any immediate effect on the power layer therefore only communication network is updated but eventually it will lead to unobservability of parts of the power layer even if does not lead to direct failure of other entities. However, this case is not considered in the simulation as it is assumed that by that time, either that communication entity will be replaced or repaired.

Identification of the K-most vulnerable entities using this setup is done by failing K-entities in the network at a time and observing the corresponding number of failed entities. Since, this process is repeated for all different combinations of K-entities in the network to verify the results obtained from the ILP solutions using MIIM [1] and IIM [2] IDRs, a small smart grid system of only 14-Buses is considered here.

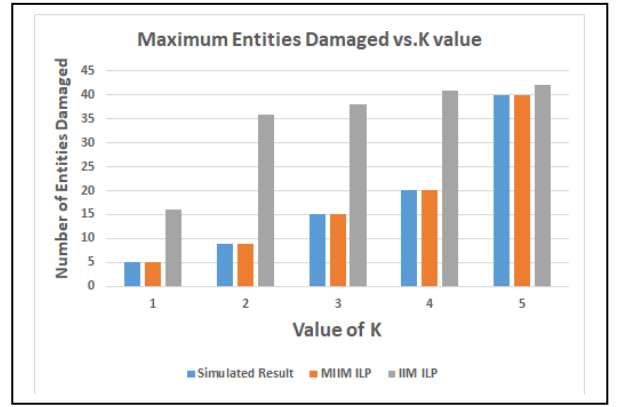

Fig. 1. Maximum total entities damaged vs. K value

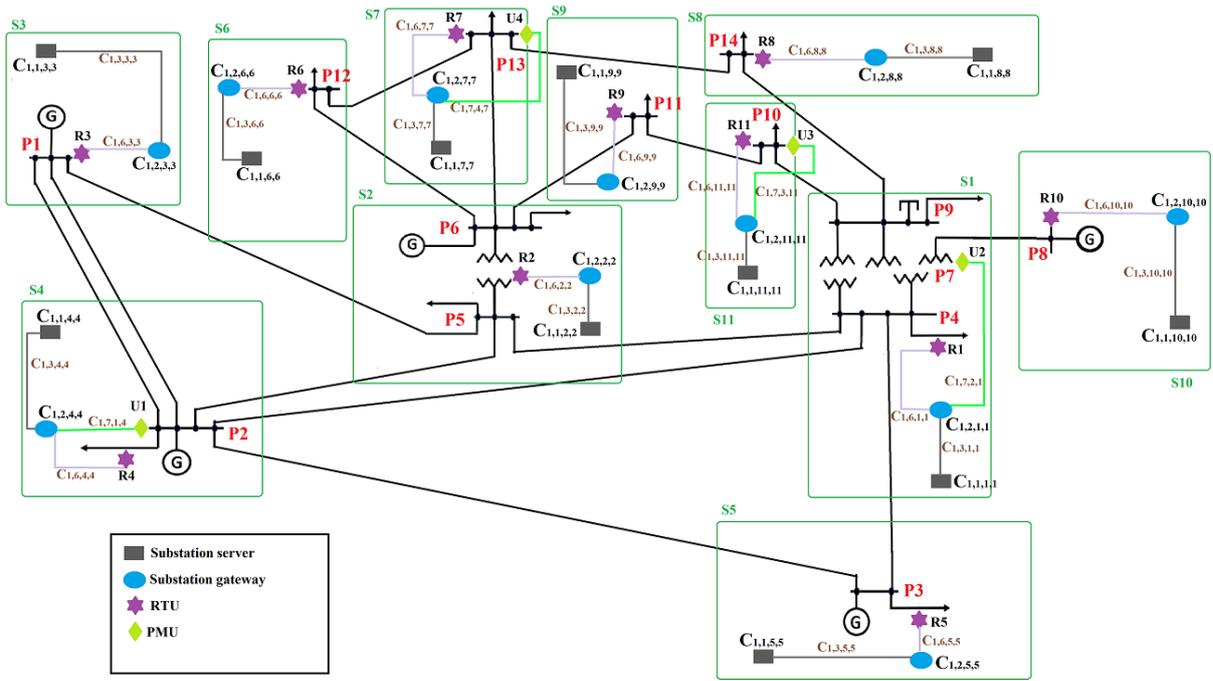

Fig. 2. Power Entities (P) and Substation Entities (SE) of a smart grid of IEEE 14-Bus system

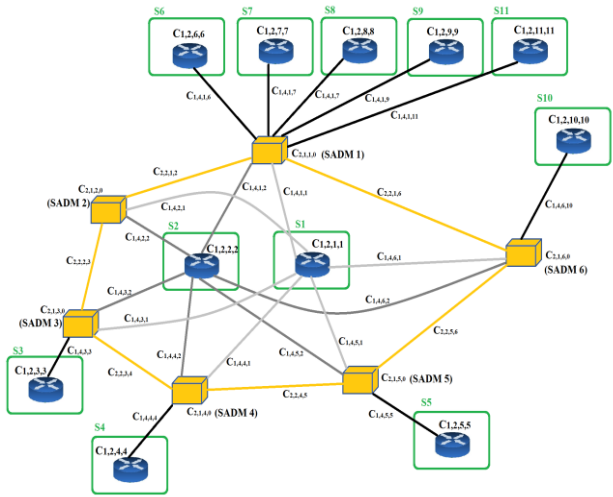

Fig. 3. SONET-Ring Entities (SRE)

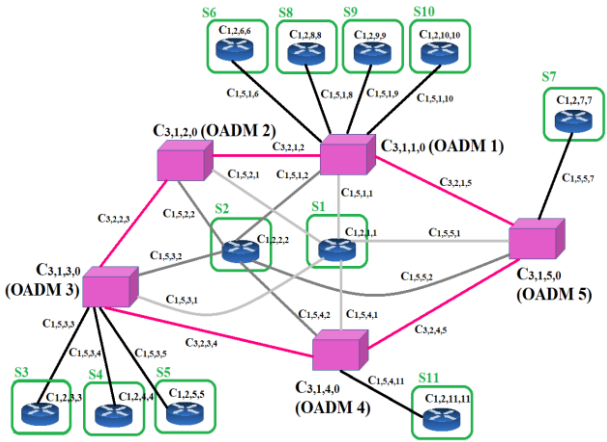

Fig. 4. DWDM-Ring Entities (DRE)

In fig.1., the maximum damage to the network after the failing of K-most vulnerable entities are predicted by the ILP based solution to the problem using MIIM IDRs and IIM IDRs. The predicted damages are compared with the simulated results for a smart grid system of IEEE-14Bus.

In the smart grid of IEEE 14-Bus system, there are 14 buses and 34 communication terminals like servers, gateways, SADMs and OADMs. It is considered that the transmission lines and communication channels can fail when the entities at the two ends of it also fail. Therefore, IDRs of those entities are considered. They can either have a state value 1 indicating they are operational or 0 denoting they have failed. However, the other 48 entities (14 *P type* and 34 *C type*) may depend on these transmission lines or communication channels and thus they are included in the IDRs of those entities. Therefore, while finding the K-most vulnerable entities, only 48 entities are taken into account, but those 48 entities also cover the other entities which belong to categories like transmission lines or communication links.

In fig.2., fig.3. and fig.4. the power entities and substation entities of communication network, the SONET-Ring entities (SRE) and the DWDM-Ring entities (DRE) are shown respectively. The nomenclature followed in the figures is same as in [1].

It is observed that for $K = 1$, the maximum damage in the network predicted by MIIM IDRs is 5 out of total 48 entities in the smart grid. According the ILP based solution using MIIM IDRs, the most vulnerable entity in the network is Bus 7 denoted as $P_7$. Failure of $P_7$ leads to the failure of $P_8$ since it is connected to $P_7$ only. As the result, the communication entities in the substation of bus $P_8$ (substation 10 or $S_{10}$) also fail. Those entities are: $S_{10}$ server denoted as $(C_{1,1,10,10})$ and $S_{10}$ gateway denoted as $(C_{1,2,10,10})$. The $SADM_6$ or $(C_{2,1,6,0})$ is connected to $S_{10}$ only, therefore it also fails. The simulation results also prove this. After the failure of $P_7$, $S_{10}$ got islanded from the rest of the network; since all entities of that substation do not contribute to the smart grid anymore, they can be considered as non-operational. This observation can propose a theory that if the smart grid system is considered as a two layered graph, then the vertices in the power layer that are connected to some pendant vertices are the most critical for $K = 1$. However, if the IIM IDRs are used, the failure of any one of the following entities:

$$\begin{Bmatrix} \{C_{1,1,6,6}\}, \{C_{1,2,6,6}\}, \{C_{1,1,7,7}\}, \{C_{1,2,7,7}\}, \{C_{1,1,8,8}\}, \{C_{1,2,8,8}\}, \\ \{C_{1,1,9,9}\}, \{C_{1,2,9,9}\}, \{C_{1,1,11,11}\}, \{C_{1,2,11,11}\} \end{Bmatrix}$$

will lead to the failure of $SADM_1$ and therefore all entities connected to $SADM_1$ or $(C_{2,1,1,0})$ will fail. This again will lead to the failure of $OADM_1, OADM_4$ and $OADM_5$. Also, due to the failure of $OADM_1$ or $(C_{3,1,1,0})$, communication entities of $S_{10}$ also fail, leading to the failure of $SADM_6$. The cascade of failure is shown below assuming substation server 6 or $(C_{1,1,6,6})$ failed initially:

$\{C_{1,1,6,6}\} \rightarrow \{C_{1,2,6,6}\} \rightarrow \{C_{2,1,1,0}\}$
$\rightarrow \{\{C_{1,2,7,7}\}, \{C_{1,2,8,8}\}, \{C_{1,2,9,9}\}, \{C_{1,2,11,11}\}\}$
$\rightarrow \{\{C_{1,1,7,7}\}, \{C_{1,1,8,8}\}, \{C_{1,1,9,9}\}, \{C_{1,1,11,11}\}\}$
$\rightarrow \{\{C_{3,1,1,0}\}, \{C_{3,1,4,0}\}, \{C_{3,1,5,0}\}\} \rightarrow \{C_{1,2,10,10}\} \rightarrow \{C_{1,1,10,10}\}$

Therefore, a total of 16 entities are damaged. This prediction does not match with the simulation results.

For, $K = 2$, according to the MIIM IDRs, the failure of both the control center servers or failure of both control center gateways can result in the maximum damage i.e. all 34 communication entities will fail. However, this situation is absurd. Moreover, the main control center is already hardened by means of the backup control center. Analyzing the effect of failure of the backup entities as well (that also in the first step), is beyond the scope of this work and also it is hypothetical. Therefore, we are considering the maximum damage caused by the next 2-most critical entities in the network. The following pairs of entities are most critical if MIIM IDRs are used:

$$\begin{Bmatrix} \{P_1, P_7\}, \{P_2, P_4\}, \{P_2, P_5\}, \{P_2, P_7\}, \{P_3, P_7\}, \{C_{1,1,2,2}, P_7\}, \\ \{P_7, P_{10}\}, \{P_7, P_{13}\}, \{C_{1,2,2,2}, P_7\} \end{Bmatrix}$$

The failure of any of the above pairs can lead to the failure of a total of 9 entities in the network. The simulation results also validate this prediction. The cascading failure of entities after the initial failure of each of the above ($K = 2$) sets is shown below:

Set 1: $\{P_1, P_7\} \rightarrow \{P_1, P_7, P_8\}$
$\rightarrow \{\{C_{1,2,3,3}\}, \{C_{1,1,3,3}\}, \{C_{1,2,10,10}\}, \{C_{1,1,10,10}\}\}$
$\rightarrow \{\{C_{2,1,3,0}\}, \{C_{2,1,6,0}\}\} = 9\ Entities$

Set 2: $\{P_2, P_4\} \rightarrow \{P_2, P_4, P_3\}$
$\rightarrow \{\{C_{1,2,4,4}\}, \{C_{1,1,4,4}\}, \{C_{1,2,5,5}\}, \{C_{1,1,5,5}\}\}$
$\rightarrow \{\{C_{2,1,4,0}\}, \{C_{2,1,5,0}\}\} = 9\ Entities$

Set 3: $\{P_2, P_5\} \rightarrow \{P_2, P_5, P_1\}$
$\rightarrow \{\{C_{1,2,4,4}\}, \{C_{1,1,4,4}\}, \{C_{1,2,3,3}\}, \{C_{1,1,3,3}\}\}$
$\rightarrow \{\{C_{2,1,3,0}\}, \{C_{2,1,4,0}\}\} = 9\ Entities$

Set 4: $\{P_2, P_7\} \rightarrow \{P_2, P_7, P_8\}$
$\rightarrow \{\{C_{1,2,4,4}\}, \{C_{1,1,4,4}\}, \{C_{1,2,10,10}\}, \{C_{1,1,10,10}\}\}$
$\rightarrow \{\{C_{2,1,4,0}\}, \{C_{2,1,6,0}\}\} = 9\ Entities$

Set 5: $\{P_3, P_7\} \rightarrow \{P_3, P_7, P_8\}$
$\rightarrow \{\{C_{1,2,5,5}\}, \{C_{1,1,5,5}\}, \{C_{1,2,10,10}\}, \{C_{1,1,10,10}\}\}$
$\rightarrow \{\{C_{2,1,5,0}\}, \{C_{2,1,6,0}\}\} = 9\ Entities$

Set 6: $\{C_{1,1,2,2}, P_7\} \rightarrow \{\{C_{1,1,2,2}\}, \{P_7\}, \{C_{1,2,2,2}\}, \{P_8\}\}$
$\rightarrow \{\{C_{1,2,10,10}\}, \{C_{1,1,10,10}\}\}$
$\rightarrow \{\{C_{2,1,2,0}\}, \{C_{2,1,6,0}\}, \{C_{3,1,2,0}\}\}$
$= 9\ Entities$

Set 7: $\{P_7, P_{10}\} \rightarrow \{P_7, P_{10}, P_8\}$
$\rightarrow \{\{C_{1,2,10,10}\}, \{C_{1,1,10,10}\}, \{C_{1,2,11,11}\}, \{C_{1,1,11,11}\}\}$
$\rightarrow \{\{C_{3,1,4,0}\}, \{C_{2,1,6,0}\}\} = 9\ Entities$

Set 8: $\{P_7, P_{13}\} \rightarrow \{P_7, P_{13}, P_8\}$
$\rightarrow \{\{C_{1,2,10,10}\}, \{C_{1,1,10,10}\}, \{C_{1,2,7,7}\}, \{C_{1,1,7,7}\}\}$
$\rightarrow \{\{C_{3,1,5,0}\}, \{C_{2,1,6,0}\}\} = 9\ Entities$

Set 9: $\{C_{1,2,2,2}, P_7\} \rightarrow \{\{C_{1,2,2,2}\}, \{P_7\}, \{C_{1,1,2,2}\}, \{P_8\}\}$
$\rightarrow \{\{C_{1,2,10,10}\}, \{C_{1,1,10,10}\}\}$
$\rightarrow \{\{C_{2,1,2,0}\}, \{C_{2,1,6,0}\}, \{C_{3,1,2,0}\}\}$
$= 9\ Entities$

When IIM IDRs are used, the following pairs become most critical entities:

$$\begin{Bmatrix} \{P_{12}, P_1\}, \{P_{13}, P_1\}, \{P_{14}, P_1\}, \{P_{11}, P_1\}, \{P_{10}, P_1\}, \{P_{12}, P_2\}, \\ \{P_{13}, P_2\}, \{P_{14}, P_2\}, \{P_{11}, P_2\}, \{P_{10}, P_2\}, \{P_{12}, P_3\}, \{P_{13}, P_3\}, \\ \{P_{14}, P_3\}, \{P_{11}, P_3\}, \{P_{10}, P_3\} \end{Bmatrix}$$

For each of the above pairs, the total damage will be: 36 entities. Considering the first ($K = 2$) set, the cascade will be:

$\{P_{12}, P_1\} \rightarrow \{C_{1,2,6,6}, C_{1,2,3,3}\} \rightarrow \{C_{1,1,6,6}, C_{1,1,3,3}\}$
$\rightarrow \{C_{2,1,1,0}, C_{2,1,3,0}, C_{3,1,3,0}\}$
$\rightarrow \{C_{1,2,7,7}, C_{1,2,8,8}, C_{1,2,9,9}, C_{1,2,11,11}, C_{1,2,4,4}, C_{1,2,5,5}\}$
$\rightarrow \{C_{1,1,7,7}, C_{1,1,8,8}, C_{1,1,9,9}, C_{1,1,11,11}, C_{1,1,4,4}, C_{1,1,5,5}\}$
$\rightarrow \{C_{2,1,4,0}, C_{2,1,5,0}, C_{3,1,1,0}, C_{3,1,4,0}, C_{3,1,5,0}\}$
$\rightarrow \{C_{1,2,10,10}, C_{1,1,10,10}, C_{3,1,2,0}, C_{2,1,1,1}, C_{1,1,1,1}\} \rightarrow \{C_{2,1,6,0}\}$
$\rightarrow \{C_{2,1,2,0}\} \rightarrow Total\ ICT\ network\ failure$

In the similar way, prediction of MIIM based ILP solution is accurate in case of $K = 3, 4\ and\ 5$ but the prediction of IIM based solution is very different from the simulation results. It is observed that from $K = 4$ to $K = 5$, there is a sudden huge change in the total number of failed entities for both MIIM based solution and simulation result. This happens due to the only set of $K = 5$, $\{P_4, P_7, P_9, P_5, P_6\}$ that leads to total communication failure in both the situations and as a result, the maximum total number of failed entities suddenly change from 20 ($for\ K = 4$) to 40 ($for\ K = 5$). The cascade of failure is shown below:

$$\{P_4, P_7, P_9, P_5, P_6\}$$
$$\rightarrow \{\{C_{1,1,1,1}\}, \{C_{1,2,1,1}\}, \{C_{1,1,2,2}\}, \{C_{1,2,2,2}\}, \{P_8\}\}$$

which means both the control center servers and gateways are failing leading to total communication network failure. Therefore, total entities failed = (initial 5 buses + P8 + 34 communication entities) = 40.

## VI. CONCLUSION AND FUTURE WORKS

Using a dual platform based simulation of the smart grid system to verify the interdependency model MIIM is a novel approach proposed in this paper. Since, the MIIM ILP based solution is verified here using this co-simulation, the K-most vulnerable entities for any huge smart grid system can be easily obtained in a short time by using the ILP based solution considering the MIIM IDRs. Therefore, this approach can be used in real smart grids to obtain a self-updating K-Contingency list just by updating IDRs after any failure in the system. A scope of future work can be finding a method to use the PMU data that is sent to the control centers at the rate of 30 samples per second to update the MIIM IDRs and automatically update the K-most vulnerable entities each time a new failure takes place in the system. Computing the ILP based contingency list within a very short period of time can be challenging and thus a suitable heuristic solution that takes into account all the observations of the simulation can also be a scope of future work.